**Research Article**

# Comprehensive Classification of Web Tracking Systems: Technological Insights and Analysis

Theofanis Tasoulas,[1] Alexandros Gazis[2], ✉ Aggeliki Tsohou[1]

[1] Ionian University, Department of Informatics, Corfu, Greece
[2] Democritus University of Thrace, Department of Electrical Engineering and Computer, Greece

**Abstract**

Web tracking (WT) systems are advanced technologies used to monitor and analyze online user behavior. Initially focused on HTML and static webpages, these systems have evolved with the proliferation of IoT, edge computing, and Big Data, encompassing a broad array of interconnected devices with APIs, interfaces and computing nodes for interaction. WT systems are pivotal in technological innovation and business development, although trends like GDPR complicate data extraction and mandate transparency. Specifically, this study examines WT systems purely from a technological perspective, excluding organizational and privacy implications. A novel classification scheme based on technological architecture and principles is proposed, compared to two preexisting frameworks. The scheme categorizes WT systems into six classes, emphasizing technological mechanisms such as HTTP protocols, APIs, and user identification techniques. Additionally, a survey of over 1,000 internet users, conducted via Google Forms, explores user awareness of WT systems. Findings indicate that knowledge of WT technologies is largely unrelated to demographic factors such as age or gender but is strongly influenced by a user's background in computer science. Most users demonstrate only a basic understanding of WT tools, and this awareness does not correlate with heightened concerns about data misuse. As such, the research highlights gaps in user education about WT technologies and underscores the need for a deeper examination of their technical underpinnings. This study provides a foundation for further exploration of WT systems from multiple perspectives, contributing to advancements in classification, implementation, and user awareness.

## Introduction

Web tracking (WT) systems are technologies that automatically record data and behaviors of internet users. Their utilization by service providers is growing rapidly. WT tools can be examined from various perspectives, including company profits, privacy protection, information security, and system architecture, [1]. For companies, WT tools are invaluable for collecting data from internet users. By analyzing this data, companies apply personalization techniques and optimize marketing and advertising strategies, [2].

However, WT technologies also pose a threat to individual privacy rights. Internet users may be aware of how companies process their data, especially when it is consciously shared (e.g., through online or registration forms). Research indicates that users often lack awareness of the extent of personally identifiable information they share or who can access it—even when shared consciously. Awareness is even more limited when data is collected via WT technologies without users' knowledge. Another critical aspect is the use of security mechanisms, like cryptography, to protect against unauthorized data disclosure or modification through WT tools. Additionally, the architecture of tracking systems is crucial, as the knowledge they produce depends on how each system operates in the background, [3, 4].

Most WT research focuses on privacy concerns, highlighting the lack of understanding about the technical workings of these systems. There is also a lack of consistent classification schemes for WT systems. For example, [5, 6], introduced a classification framework based on observable behaviors and properties, such as a tracker's scope (within-site or cross-site tracking). Similarly, [6, 7, 8], developed a scheme based on how websites use WT tools. Such works are essential for better understanding WT systems. Currently, most internet users have limited familiarity with WT systems (beyond cookies) and possess only basic knowledge of these technologies, as shown by recent surveys, [5, 9, 10, 11, 12].

WT systems are web software tools that collect data and store it in files or databases. This data is transmitted through internet communication protocols, such as HTTP, during interactions between a web client (e.g., a browser) and a web server. The process begins with a request from the web client. For example, when a web client requests to view a website (i.e. the front end technologies consisting of: static HTML and CSS or dynamic webpages of JavaScript, and HTML/CSS files), it sends information through the HTTP protocol, [13].

If the client does not send its IP address, the server cannot know where to deliver the website. This information is added to the HTTP protocol in text form. After the client gathers the requested data, the server sends the website to the client and stores the data received from the client (e.g., IP address, operating system details). These data points can be saved on either the client's or the server's side.

Data can also be transferred via APIs (application programming interfaces), which are sets of subroutine definitions, communication protocols, and programming tools designed to enhance software functionality. APIs make tracking significantly easier and more efficient. There are various methods to track user data, which will be examined in the following sections, [7, 14].

# Related Work

In this section, we briefly describe the necessary technological components that are connected with web tracking technology and web tracking systems.

# Cookies

HTTP cookies are commonly used by website owners and third parties to monitor users and their behavior through the HTTP protocol. They are among the most well-known WT systems. A cookie is not a program but a text file containing key-value pairs (e.g., Visiting Time = 21:12, IP Address = 127.0.0.0). HTTP cookies are stored on the client's computer.

Each time a user visits a website, the website's server requests the cookie to easily access the necessary data. If no cookie exists for that site, the server assumes the user is visiting for the first time. The server then creates a new identifier in its database for the visitor and sends a cookie. Cookies store user data, such as passwords, allowing users to avoid reentering them each time they log in. This benefits both websites and internet users, [7, 8, 15, 16, 17, 18].

Permanent cookies, also known as zombie cookies or ever-cookies, are HTTP cookies controlled by scripts. When deleted, they are recreated from backups stored in the HTML5 local storage or on the client's hard drive. This occurs when a user visits a webpage that secretly sends a script or executable program via the browser. The script checks a specific location on the client's disk to restore the deleted cookie from a backup or save the cookie's current state, [19, 20].

## Local Shared Objects

Local shared objects are cookies that do not rely solely on the HTTP protocol for tracking but require software manipulation. For example, Flash cookies are managed by Adobe Flash Player, while Silverlight Secure Storages are handled by Microsoft Silverlight, [21, 22]. These development tools enhance web content and user experience by using their own APIs and scripts to communicate with browsers, retrieve data from viewed pages, or access browser storage. The collected data is stored in local shared objects, which may include small databases, JSON-formatted files, or complex storage structures beyond simple key-value texts. These objects improve the user experience by storing privacy preferences, restoring time points (e.g., where a user stopped playing a Flash video), and more. Additionally, they can link a user's activities across different websites or browsers and synchronize their cookies. Cookie synchronization involves linking data from various cookies using fingerprinting techniques to create user interest profiles. Local shared objects have a larger storage capacity than regular cookies and are harder to delete due to their ability to be recreated by the associated software, [5, 8, 9].

## HTML5 Tracking

HTML5 offers numerous APIs to enhance the web experience, such as server-sent events, XMLHttpRequest2, Web Messaging, and Geolocation. By using HTML5 APIs, websites or software can gather user data through the browser and store it either on the server or the client's storage. Additionally, HTML5 introduces new client-side storage mechanisms, including Web Storage, Local Storage, and Indexed DB. These options allow for more efficient data storage using JSON-formatted files,

NoSQL techniques, and API calls. Consequently, HTML5 provides an effective means of tracking through API interactions, [7, 8, 10, 14, 23].

### Web beacons

A Web beacon (also known as a web bug, tracking bug, pixel tag, or clear GIF) is a technique that involves displaying a very small (typically 1x1 pixel) image in an HTML file or web programming code (e.g., JavaScript). For the image to appear, the user's browser must send a request to the server hosting it. This request allows the server to identify which user wants to view the image.
Due to its small size, the Web beacon is usually invisible to users, though it can be detected by examining the source file. This enables tracking without users' knowledge. Additionally, the image's URL may include a script written in server-side programming languages like PHP or ASP.NET. When executed on the server, this script collects even more user data. Typically, all data gathered by a Web beacon is stored on the server, [7, 8, 15, 23].

### Spyware

Spyware tools are software programs installed on a computer to monitor activity or cause harm. They can access third-party APIs or components from browsers or operating systems (OS), such as Layered Service Providers and IConnectionPoint, to retrieve user data. Once collected, this data is saved in hidden files, and a background process transmits the files to a specific server via HTTP. Additionally, Spyware often collaborates with other software or techniques, such as Adware or keystroke loggers. Keystroke loggers operate at the kernel level, monitoring binary data sent by input devices (e.g., keyboards) through OS APIs, allowing them to capture sensitive information like passwords.
Most Spyware tools are illegal and lack authorization for download, installation, or access, [24, 25, 26].

### Email Tracking

Email tracking is implemented using a web beacon or email services. It can operate on various networks (public, local, etc.) to reveal how often a message is read or transmitted and by whom, along with their IP address. If a server collects data from a specific web beacon, it can determine which email was opened, where, and when. Modern email services use scripts (usually involving POST and GET methods) on their servers to track the communication channel, sender, and receiver. They may also use web analytics services. All this data is stored on servers, [27, 28, 29].

### Local Based Services

Recent advances in wireless location tracking, such as cellular networks or RFID tags, offer many opportunities to implement or enhance Location-Based Services. Using APIs and maps enables real-time traffic analysis, as APIs (like Google Maps API) provide extensive geographic information. APIs repeatedly communicate with telecommunication systems managed by location providers and brokers, requiring a stable connection between the client, server, and provider. Location-Based Services often process queries based on current location (e.g., "Where is the nearest hotel?") by calculating distances between the current and target locations. These services also track a person's location (e.g., via GPS) to improve traffic flow or assist with navigation to specific destinations, [30, 31].

### Fingerprinting

Fingerprinting is a technique used to create a unique string that identifies internet users. This string is stored in a cookie or browser storage, allowing websites to request it and identify visitors. Trackers use fingerprints to enhance tracking, build user behavior profiles, and provide personalized services. They can also be used for multifactor authentication and improved security. Some fingerprints are generated via HTTP mechanisms like ETags or scripts, while others are more complex, based on collected data. For instance, canvas fingerprinting at times uses the Canvas API, [32], to gather information like fonts, graphics card details, screen size, and other features, which are then hashed to produce the fingerprint. Similarly, various APIs enable methods like audio, battery-based, font fingerprinting, and more, [33, 34, 35, 36].

### Taint Tracking

Taint tracking, or information flow tracking, monitors how applications access and manipulate personal data. Specifically, when an application initiates one or more processes in IPC (Inter-process Communication), a taint tag is added to the process based on specific dataflow rules, [37]. This can be implemented using virtual or hardware components, such as a VM interpreter or shadow memory, to access low-level OS APIs. The taint tracking system labels sensitive data sources and temporarily tags them as they flow through an application. When tainted data are transmitted over a network, the system records the tags, the responsible application, and the destination. This provides real-time feedback to users about how their data is being used, [38, 39, 40].

### Web Privacy Measurement

Many researchers use Web Privacy Measurements to track tracking services on websites. These tools employ browsers and task managers to convert high-level commands (e.g., "visit a website") into specific subroutines, distribute commands to browser administrators (e.g., create command threads per browser), and interact with browser APIs to analyze data flow. They also use COM interfaces or IConnectionPoint for deeper data collection and tracker detection. Additionally, Web Privacy Measurements crawl files and scripts from websites or data collected by them, manipulating and storing this information in databases for analysis. By crawling websites, they can identify WT systems, such as detecting "src" tags in HTML files leading to web analytics services or tracking scripts (e.g., cookies, fingerprints). Advanced tools can even reveal how personal data is used. These complex systems rely on APIs, databases, browser interfaces, and scripts, and often integrate machine learning algorithms, datasets from privacy tools like Ghostery, and more, [41, 42, 43, 44].

## Other Tracking techniques

**Session IDs in Hidden Fields**

Before cookies, users were tracked through session IDs stored in hidden fields. Basic user data from the HTTP protocol (e.g., IP address) or an identification string from ETag could be passed to another web page via the URL (using the GET method) or a hidden web form field (using the POST method). While this method works without client-side programming languages (e.g., JavaScript), it is limited to a single browsing session, [45, 46].

**Web-form authentication**
Some websites restrict resources to signed-up users, requiring login via web-form authentication (e.g., username and password). This method makes user identification accurate and straightforward, allowing WT systems to record all user activity on the website, whether stored on the client or server. It is also independent of the user's web browser, operating system, computer, or location, [45, 47].

**Cookies synchronization**
Cookies synchronization or cookies syncing is a process used by websites' server, in order to transmit identifiers or other elements of the users' cookies. In this way cross-tracking is achieved, which is an effective way of sharing cookies' data and form a profile for each user. Consequently, the user's preferences and behaviors are linked from various tracking websites, [9, 16, 45, 46, 47, 48].

**Web analytics services**
If a website wants to analyze its traffic, they can add a tag on their HTML files which leads to web analytics services such as Google Analytics. Then, the scripts available in the web analytic services will run on their server, in order to collect and analyze data from visitors in real-time.

**Table 1:** Classification scheme based on technological architecture

| Class A<br>HTTP Tracking | Class B<br>API Tracking | Class C<br>User identification | Class D<br>Complex tracking |
|---|---|---|---|
| **A1 (Client storage)** | **B1 (Self owned API)** | | |
| Web Form Authentication | Flash cookies and local shared objects | Canvas fingerprinting | Analysis services |
| Session IDs in hidden fields | Persistent cookies, zombie cookies, evercookies | Synchronization of cookies | Web Privacy Measurements |
| HTTP Cookies | Local Based Services | Etags | |
| | HTML5 | | |
| **A2 (Server storage)** | **B2 (Third party API)** | | |
| HTML | Taint tracking | | |
| Web Beacons | Spyware | | |
| Email Tracking | | | |

User's data are saved in both client's and service's storage. As long as those services run on different websites, the service can define cookies in the user's browser, which will contain a unique identifier. As a result, user will contain same cookie identifiers in lots of different websites cookies, making cookie synchronization very effective and yet very simple process. Web analytics services depend on big databases and servers for great manipulation of users' data in real time. Thus, data are saved in both client's and website server's storage, mostly on servers, [9, 16, 49].

# Materials and Methods

## Materials

The proposed classifying tracking systems is not very common in the literature, even though it can provide a better understanding on those systems. A classification scheme allows the separation of certain objects into categories, according on specified criteria. The current paper investigates WT systems based on how they work and their technological architecture and for that reason the proposed classification is developed based upon this. It's clear that HTTP and APIs are the main methods of tracking, either if the data are stored on client either on server. Moreover, fingerprints are usually created by APIs or scripts, but it would be better to separate them from any other WT system. That's because fingerprints don't collect data, but it's an exclusive category of identifying internet users in order to help other tracking mechanism. And there are more complicated tracking systems as well, combining all the previous tracking methods. Hence the current scheme proposed 6 classes, which are represented in Table 1. Lastly to mention, this scheme is not based on any other previous research effort.

### Class A – HTTP Tracking

This class includes technologies that rely on the HTTP protocol to collect data, often using HREFs, URLs, or HTTP GET requests. Code must be executed to store the data in files or databases, which can run either on the server or client. The data storage location isn't necessarily tied to where the code is executed. For example, code may run on the server, while data is stored on the client. Users can interact with these technologies by analyzing executable code and HTML files. Most tracking technologies depend on HTTP, as it facilitates basic communication between the server and client. However, this class excludes more complex, combinational methods. Therefore, Class A includes the following technologies:
  a) Web Form Authentication
  b) Session IDs in hidden fields
  c) HTTP Cookies
  d) HTML
  e) Web Beacons
  f) Email Tracking

Class A can be further divided into A1 and A2. A1 consists of technologies that store data on the client's computer, while A2 refers to technologies that store data directly on their own servers. Thus, a), b), c) belong to A1, while d), e), f) are assigned to A2.

### Class B – API Tracking

Technologies in Class B collect and store data in storage structures (databases or files) using APIs or other interfaces that mediate communication between objects and their environment. These interfaces often rely on low-level operating system APIs (such as Layered Service Providers), application APIs (e.g., HTML5, Adobe Flash), and browser interfaces (e.g., IConnectionPoint, COM). Users interact with these systems through web platforms, applications, or web pages. Implementing these tracking systems requires significant coding and a deep understanding of how APIs function. In conclusion, Class B includes the following technologies:
  a) Flash cookies and local shared objects
  b) Persistent cookies, zombie cookies, evercookies
  c) Local Based Services
  d) HTML5
  e) Taint tracking
  f) Spyware

Class B can be divided into B1 and B2. B1 includes tracking systems that use their own software APIs, while B2 comprises WT systems that rely on third-party APIs from programs, operating systems, or browser interfaces. Thus, a), b), c), d) belong to B1, while e), f) belong to B2.

### Class C – User Identification

Class C includes WT techniques that identify specific internet users. These systems collect data to create fingerprints, which uniquely identify users. A script or API generates a fingerprint based on collected data, storing it in cookies, local shared objects, or other storage. This process can be executed via simple scripts or APIs. The unique identifiers distinguish users as visitors to web pages or web browsers. Users may encounter these identifiers in cookies or local shared objects, often encrypted and difficult to recognize. Therefore, Class C includes the following systems:
  a) Canvas fingerprinting
  b) Synchronization of cookies
  c) E-tags

### Class D – Complex tracking

Other tracking technologies combine HTTP, APIs, and web technologies or programming applications for more complex tracking. These systems often rely on large servers and databases to function effectively, as they may operate across multiple websites simultaneously, requiring significant resources. Additionally, complex WT systems often have research or commercial contexts that remain undisclosed, making them less known to both internet users and researchers. Lastly, Class D includes the following systems:
  a) Analysis services
  b) Web Privacy Measurements

**Clarification on Categorization Criteria**

It is noted that the classification of each tracking technology into a specific category was based on its core mechanism, i.e., its operational properties and the generic technological layering it relies on. For instance, session IDs in hidden fields are placed under Class A (HTTP Tracking) because they operate through strictly HTTP protocol and do not require any scripting or external software entities. In contrast, persistent cookies do not operate purely through HTTP protocols, as, for example, zombie cookies or evercookies are included in Class B (API Tracking) as they depend on browser-based APIs or even client-side scripts (e.g., Flash or HTML4) to recreate or manipulate the stored data in order to extend their functionality beyond the strict boundaries of HTTP behavior. As such, this distinction reflects whether a system may function using straightforward protocol-level interactions or it may employ a more complex interface and programming logic for tracking.

## Methods

Based on the four classes analyzed above, Table **1** showcases a comparison of these different classes. In this section, we compare our work with two previous studies. In [5], the authors introduced a classification framework, classifying web trackers based on tracking behaviors and properties. Specifically, the scheme identifies 5 behaviors and 7 properties, primarily observable from the client's side. A web tracker is classified into a behavior if it exhibits at least one of the specified properties, as detailed in Table **2** and Table **3**, [5].

Another classification scheme was proposed in Imane Fouad's research to identify tracking domains and their interconnections. The authors in [7], focused on analyzing WT webpages using web beacons (pixel tags) by detecting these beacons. The arrows in [7], representing potential relationships between categories in a stateful crawl, suggest that categories within each class likely interact sequentially, as shown in Table **4**.

By crawling 829,349 webpages and detecting web beacons, a tracking classification framework was developed, categorizing web trackers based on their behavior. This scheme consists of 4 classes and 7 categories, each explaining how the WT domain operates, its impact on user privacy, and statistical results from the research dataset. A web tracker using web beacons is classified based on its tracking behavior, as outlined in [7].

The Franziska Roesner classification scheme (Table **2** and Table **3**) focuses on web tracker behaviors and properties, primarily high-level criteria observable from the client (user's browser). In contrast, our scheme (Table 1) relies on lower-level criteria, emphasizing programming implementation and technological background, mostly observable from the server's side. Additionally, Imane Fouad's scheme (Table **4**) covers more complex, non-basic WT systems based on the perspective of website usage. Our classification (Table **1**), however, includes simpler mechanisms like HTTP and session IDs in hidden fields, classifying software rather than the website's tracking methods. In conclusion, the differences between these classification schemes stem from differing research perspectives, providing a more comprehensive understanding of WT mechanisms, [5, 7].

The next section will discuss internet users' awareness and their perspective on WT systems. The classification of WT systems in this research (Table **1**) has limited relevance to users' perspective, as it focuses on observable behaviors from the browser rather than what occurs on the server side. For example, users may notice cookie notifications but are less aware of fingerprints. Consequently, the most recognized WT tools are cookies (96.9% familiarity) and GPS (68.9%), while other tools are 30% less well-known (Appendix, Questions 7, 18). In conclusion, previous research and classification efforts are more aligned with users' perspective, which relies on observable behaviors from the client side. In contrast, this paper's classification, focused on server-side criteria, is less related to users' awareness.

## Results and Discussion

The results are twofold. Firstly, we specify the questionnaire structure, followed by an explanation of the study's metrics and statistics.

It is noted that, to gain a detailed understanding of the user's knowledge and perceptions so as to analyze and explain what web tracking systems mean to them, the questionnaire was carefully designed with both structure and accessibility in mind. Specifically, it consisted of clear, concise questions formulated to avoid technical jargon or other misconceptions so as to ensure that participants, even without a strict technical or computer science-related background, could respond in a meaningful manner. As such, the survey included demographic items followed by targeted questions assessing both factual knowledge (identification of cookies/tracking methods) and subjective awareness (e.g., perceived danger or familiarity with terms). Multiple-choice questions were balanced with open-ended options, allowing for both quantitative future analysis and richer user insights. Our aim was not to merely explore what users know, but also to try to understand how they interpret and react to WT technologies.

**Table 2:** Classification scheme based on observable behaviors, tracking categories as derived from [5]

| Category | Name | Profile Scope | Description | Example | Visit Directly? |
|---|---|---|---|---|---|
| A | Analytics | Within-Site | Acts as a third-party analytics engine for individual websites. | Google Analytics | No |
| B | Vanilla | Cross-Site | Employs third-party storage to track users across multiple websites. | Doubleclick | No |
| C | Forced | Cross-Site | Strictly requires users to visit the tracker directly (e.g., through popups or redirects). | InsightExpress | Yes (forced) |
| D | Referred | Cross-Site | Relies on another tracker (B, C or E) to establish unique identifiers. | Invite Media | No |
| E | Personal | Cross-Site | Accessed directly by users in other contexts. | Facebook | Yes |

**Table 3:** Classification scheme based on observable behaviors, tracking begaviour as mechanism as derived from [5]

| Property | Behavior |
|---|---|
| Tracker creates state owned by the site itself (first-party state). | A |
| Requests made to the tracker expose site-owned data. | A |
| The third-party request to the tracker contains state from the tracker. | B, C, E |
| Tracker establishes its state from a third-party position; users do not directly engage with the tracker. | B |
| The tracker forces users to visit it directly. | C |
| Relies on interactions with another tracker (A, B, C, or E) to transmit data (not originating from the site itself). | D |
| Users voluntarily access the tracker's site. | E |

**Table 4:** A classification of tracking methodologies, grouping techniques into classes and categories based on functionality and mechanisms, [7]

| Class | Category | Description |
|---|---|---|
| **Explicit Cross-Domain Tracking** | Basic Tracking | Standard tracking functionality, including the integration of third-party trackers. |
| | Third Party Included by a Tracker | Third-party domains directly included by the tracker for data collection. |
| **Implicit Cross-Domain Tracking** | Basic Tracking Initiated by a Tracker | Tracking initiated indirectly through other mechanisms. |
| | Third Parties That Include Trackers | Third parties that themselves host or incorporate tracking scripts. |
| **Cookie Syncing** | First to Third Party Cookie Syncing | Syncing cookies from first-party domains to third-party trackers. |
| | Third to Third Party Cookie Syncing | Syncing cookies between multiple third-party trackers. |
| | Third Party Cookie Forwarding | Forwarding cookies between third parties for enhanced tracking. |
| **Analytics** | Analytics | Analytical tools providing site-specific or cross-domain insights. |

## Questionnaire structure

In this research, an online questionnaire survey was conducted using Google Forms, with 1,032 participants from Greece. Different types of surveys offer varying advantages and disadvantages depending on the research goal. For our study, we chose an online questionnaire via Google Forms to reach as many internet users as possible and ensure ease of data collection. This approach was simple to create and share online, guaranteeing anonymity and no time pressure for participants (allowing them to review questions as needed). The questionnaire aimed to explore how much internet users know about WT systems and the factors influencing their knowledge. Additionally, we examined whether a lack of awareness about WT systems affects users' privacy concerns.

To ensure the questions were understandable for all participants, regardless of their technical knowledge, we avoided jargon and technical terms, opting for small, closed questions (up to 21 words). Most questions included open-ended answers, such as "other, describe in a few words." The data was analyzed using distributional descriptions, statistically comparing groups (e.g., correct vs. incorrect answers) and identifying possible associations between variables. The questionnaire began with demographic questions, including gender, age, daily internet usage, and computer science background (Appendix, Questions 1-4) to identify factors influencing awareness of WT systems. It then included four multiple-choice questions on WT systems, each with three incorrect answers and one correct option, such as: "Do you know what cookies are? a) Text files b) Software c) Virus d) I don't know" (Appendix, Question 8). Some multiple-choice questions were more subjective or debatable, such as: "Which of the following tracking systems do you know? a) Local-based service b) Flash Cookies c) …" (Appendix, Question 18), [50, 51, 52].

Lastly, regarding the survey methodology used, it is noted that participants were reached online using anonymous sharing, i.e., sharing of Google Forms; thus, we aimed for wide accessibility and voluntary participation. Specifically, we paid extra attention to ensure that the questionnaire was intentionally simple so as to avoid bias from technical terminology, and closed-ended questions were used as a means to facilitate statistical comparison. However, it must be noted that the sampling method may introduce limitations via this mode of operation, as it may lead to self-selection bias, and thus the sample may not fully represent the broader spectrum of the population. As such, while descriptive statistics and basic comparison were used as a means to analyze responses, in future analysis of a bigger sample, it should also help to include regression models to quantify the impact of different variables on user awareness and concern.

**Structure Analysis**

The first objective was to uncover how much internet users know about tracking systems. The average correct answers across four multiple-choice questions was 37.63%, although only 26.4% claimed to understand how WT systems work (Appendix, Question 6). People appear aware of the data websites collect and the reasons behind it, yet only 25.2% are familiar with the WT systems (33.6% including cookies) that websites employ (Appendix, Questions 16-19). In conclusion, internet users have some knowledge about tracking systems, but at a very basic level.

The second goal of the questionnaire was to identify the most significant factor affecting internet users' awareness (gender, age, daily internet usage, and relation to computer science). To achieve this, queries in SQL were utilized within an Oracle Apex database. The sum of correct answers was calculated for each factor. For example, if 51% of men answered correctly to Question 8, and 54% answered correctly to Question 9, the sum for men would be 51% + 54% + ... The goal was to assess the deviation between lower and higher correct answers for each factor. For the "Age" factor, 13-18-year-olds had the lowest sum of correct answers (446%), while 19-24-year-olds had the highest sum (540%). This indicates that younger people know the least, whereas 19-24-year-olds know the most about WT systems. The deviation is the difference between these percentages, 540% - 446% = 94%, indicating that age affects awareness by 0.94. Similarly, gender affects awareness by 0.78, daily internet usage by 0.7, and relation to computer science by 1.42. These numbers provide a comparative measure of how each factor influences correct answers. Thus, the "relation to computer science" emerged as the most important factor affecting internet users' awareness of WT systems.

The third and final objective was to examine whether internet users' lack of awareness correlates with increased concerns about WT tools. To accomplish this, SQL queries were used to identify users who answered with the term "virus" (i.e., "cookies are viruses" from Appendix, Question 8). These users (203 in total) are assumed to be concerned about WT systems. The average of correct answers for those concerned is 0.39 (let A1). By repeating this process for users not concerned about WT systems, the average of correct answers is 0.37 (let A2). If A1 significantly differed from A2, it would suggest that concerned users have less knowledge about WT systems than non-concerned users, implying that awareness reduces concerns about WT systems. However, since A1 and A2 do not differ significantly, it is concluded that awareness about WT systems is independent of concerns regarding WT tools. While initial results showed that awareness does not significantly correlate with concern about data misuse (as the average correct answers between concerned and unconcerned users were 0.39 and 0.37 respectively), further statistical analysis could help validate this finding. Future work could involve the use of inferential statistical methods such as hypothesis testing or regression models to confirm the absence or presence of statistically significant differences between user groups. This would provide a more robust understanding of whether concern is indeed independent of awareness or influenced by other hidden variables.

# Conclusions

This paper provides a deeper understanding of the technological background of existing WT systems, which is largely underexplored in the literature. Other studies often examine aspects such as privacy impact, but they rarely analyze how these systems function. The current research focuses on the technological foundation of WT systems and introduces a new classification system based on their technical workings. We then compare our classification with prior systems. Other classification schemes can be developed depending on the perspective from which WT systems are studied (e.g., privacy impact, security, algorithmic efficiency), leading to more comprehensive knowledge of these systems. Furthermore, there may be other WT systems not covered by this study, especially those with research or commercial contexts. A deeper investigation into these systems will foster the development of more efficient WT technologies across various domains.

This paper also examines internet users' knowledge of WT systems through an online survey, identifying the key factors influencing their awareness. We found that "relation to computer science" is the most significant factor affecting knowledge, surpassing other variables like gender, age, and daily internet usage. Our findings reveal that internet users' understanding of WT systems remains at a very basic level. Additionally, the study shows that a user's lack of awareness doesn't necessarily translate into heightened concerns about WT systems. However, the paper does not address how users can acquire a better understanding of these technologies. As WT systems continue to evolve, finding effective ways to increase awareness—especially among those not familiar with computer science—remains essential.

Although this study focuses on the technological classification of web tracking systems and excludes the organizational and privacy implications by design, it is important to note that many of the tracking methods analyzed may pose significant security and privacy risks. Specifically, we must acknowledge that, techniques such as persistent cookies, fingerprinting, spyware, and cookie synchronization can be used to track users across websites without explicit consent, thus potentially providing a gateway for unauthorized profiling or data exploitation. As such, while these concerns were not the central aim of this research, future studies should expand on the proposed classification scheme by integrating risk assessment or evaluating the ethical implications of the technological categories.

Lastly, it is also acknowledged that several WT systems used in this research, or commercial environments may not be fully covered in this study. Analytically, although the classification is based on widely used and technically established tracking methods, emerging technologies such as AI-enhanced fingerprinting techniques or blockchain-based tracking mitigation strategies, and specifically tracking mechanisms embedded in IoT ecosystems, represent promising areas for future analysis. As such, including such systems could provide a broader classification that would strengthen the taxonomy and categorization, but also make the study more adaptable to current and evolving digital environments. Additionally, understanding the technical characteristics of such systems can also support regulatory efforts, such as improving GDPR compliance mechanisms. This is particularly important, as it helps scientists define clearer standards for what constitutes tracking and ensures that new technologies are addressed by data protection frameworks.

# Author Contributions

The authors equally contributed in the present research, at all stages from the formulation of the problem to the final findings and solution.

# Conflict of Interest

The authors have no conflicts of interest to declare that are relevant to the content of this article

# Funding

No external funding was received for this research.

# Availability of Data and Materials

Data supporting the results of this study are available upon request from the corresponding author.

# Author Contribution

Validation, Writing—review, Data curation, Conceptualization, Project administration, Supervision: Aggeliki Tsohou

Methodology, Formal analysis, Investigation, Writing, Citation —original draft preparation: Theofanis Tasoulas

Validation, Writing—review and editing, Resources, Citation editing, Template Formatting: Alexandros Gazis

# Appendix

## List of Abbreviations

| Abbreviation | Full Term |
| --- | --- |
| WT | Web Tracking |
| IoT | Internet of Things |
| API | Application Programming Interface |
| HTTP | HyperText Transfer Protocol |
| HTML | HyperText Markup Language |
| CSS | Cascading Style Sheets |
| OS | Operating System |
| GPS | Global Positioning System |
| COM | Component Object Model |
| VM | Virtual Machine |
| IPC | Inter-process Communication |
| DB | Database |
| SQL | Structured Query Language |
| ID | Identifier |
| NoSQL | Not Only SQL |
| GET | HTTP GET method |
| POST | HTTP POST method |
| GDPR | General Data Protection Regulation |
| CSS | Cascading Style Sheets |
| HTML5 | HyperText Markup Language version 5 |
| JSON | JavaScript Object Notation |

## Survey Questions

In this section we showcase the Survey's Questions:

1) What gender are you?

☐ Man
☐ Woman

2) What is your age?

☐ 0 to 12
☐ 13 to 18
☐ 19 to 24
☐ 25 to 38
☐ 39 to 55
☐ 56 and above

3) How much time do you spend online?

☐ I rarely use the internet
☐ I use the internet sometimes
☐ 0 to 2 hours a day
☐ 2 to 5 hours a day
☐ More than 5 hours a day

4) What is your relation with Computer Science?

☐ I have not studied Computer Science
☐ I have not studied Computer Science, but my profession is related to it (e.g. I do some programming)
☐ I currently study Computer Science at University or college
☐ I have a degree (bachelor, master degree, etc.) in Computer Science

5) Do you know that the websites you visit online record data about you? (e.g. When did you visit a web site and from which device (smartphone, computer, etc.)?

☐ Yes
☐ No

6) Do you know how this is done?

☐ Yes, I know
☐ Yes, but just a bit
☐ No, I don't know

7) Have you ever heard of cookies?

☐ Yes
☐ No

8) What do you think cookies are?

☐ They are text files
☐ They are software
☐ They are viruses
☐ I don't know
☐ Other (please describe in a few words)

9) Do you think cookies are dangerous?

☐ Yes
☐ Probably
☐ No, I don't think so
☐ I don't know

10) Do you think a website can collect your information if you just view a picture online?

☐ Yes
☐ Probably
☐ No, I don't think so
☐ I don't know

11) Do you think a website can collect information about you through HTML (without cookies)?

☐ Yes
☐ Probably
☐ No, I don't think so
☐ I don't know

12) Do you think two or more sites can share cookies?

☐ Yes
☐ Probably
☐ No, I don't think so
☐ I don't know

13) Do you think it's easy for a website to distinguish two or more people? (e.g. to distinguish George from Maria)

☐ Yes
☐ Probably
☐ No, I don't think so
☐ I don't know

14) Do you think there are ways to prevent a site from tracking you?

☐ Yes

☐ Probably
☐ No, I don't think so
☐ I don't know

15) Do you think Ad blocker or incognito mode makes it difficult for a site to track you?

☐ Yes
☐ Probably
☐ No, I don't think so
☐ I don't know

16) Do you think it's possible to track you when they send you an email? If so, how? (select one or more answers)

☐ Yes. They use web software (e.g. in JavaScript).
☐ Yes. They use viruses.
☐ Yes. They use images.
☐ Probably
☐ No, I don't think so
☐ I don't know

17) What information do you think a site can collect from you? (select one or more answers)

☐ Date and time I visited a site
☐ Browsing history (previous visits to other sites)
☐ Location (e.g. city, home address)
☐ Browser (e.g., Google Chrome, Mozilla Firefox)
☐ Operating system (e.g., Windows, IOS, Android)
☐ IP address
☐ Data from other cookies on my computer
☐ Passwords
☐ Credit card numbers
☐ Files from my computer (e.g., photos, videos, music)
☐ I don't know

18) Do you know any of the following tracking systems? (select one or more answers)

☐ GPS or other Local Based Service
☐ Flash cookies
☐ Spyware
☐ Email tracking
☐ Web beacon (or Web bug, tracking bug, pixel tag)
☐ HTML5 Tracking
☐ Taint Tracking
☐ I don't know any of them

19) Finally, why do you think websites collect your personal data? (select one or more answers)

☐ Maybe they want to hurt me
☐ For security reasons and to know that I am not doing anything illegal on the internet
☐ To improve their services
☐ To make money, mainly through advertising
☐ To sell my information to third parties and make money
☐ I don't know
☐ Other (please describe in a few words)